\pdfoutput=1

\documentclass[11pt,twoside,a4paper,cmspaper,final,collab]{cms-tdr}
\def\svnVersion{326805}\def\cmsCernNo{2013-037}\def\cmsCernDate{\today}\def\cmsMessage{Published in Physical Review Letters as \href{http://dx.doi.org/10.1103/PhysRevLett.116.071801}{\doi{10.1103/PhysRevLett.116.071801.}}}

\begin{document}\cmsNoteHeader{EXO-15-001}

\hyphenation{had-ron-i-za-tion}
\hyphenation{cal-or-i-me-ter}
\hyphenation{de-vices}
\RCS$HeadURL: svn+ssh://svn.cern.ch/reps/tdr2/papers/EXO-15-001/trunk/EXO-15-001.tex $
\RCS$Id: EXO-15-001.tex 317801 2016-01-15 12:12:05Z alverson $
\newlength\cmsFigWidth
\ifthenelse{\boolean{cms@external}}{\setlength\cmsFigWidth{0.98\columnwidth}}{\setlength\cmsFigWidth{0.75\textwidth}}
\ifthenelse{\boolean{cms@external}}{\providecommand{\cmsLeft}{top}}{\providecommand{\cmsLeft}{left}}
\ifthenelse{\boolean{cms@external}}{\providecommand{\cmsRight}{bottom}}{\providecommand{\cmsRight}{right}}
\ifthenelse{\boolean{cms@external}}{\providecommand{\cmsLeft}{top}}{\providecommand{\cmsLeft}{left}}
\ifthenelse{\boolean{cms@external}}{\providecommand{\CL}{C.L.\xspace}}{\providecommand{\CL}{CL\xspace}}
\ifthenelse{\boolean{cms@external}}{\providecommand{\CLp}{C.L}}{\providecommand{\CLp}{CL}}
\ifthenelse{\boolean{cms@external}}{\providecommand{\NA}{\ensuremath{\cdots}\xspace}}{\providecommand{\NA}{\text{---}\xspace}}

\cmsNoteHeader{EXO-15-001}
\title{Search for narrow resonances decaying to dijets in proton-proton collisions at \texorpdfstring{$\sqrt{s}=13$\TeV}{sqrt(s) = 13 TeV}}

\newcommand{\mjj}{\ensuremath{m_{\text{jj}}}\xspace}
\newcommand{\mjjFAT}{\ensuremath{m^\text{Wide}_{\text{jj}}}\xspace}
\newcommand{\dphijj}{\ensuremath{\abs{\Delta\phi{\text{jj}}}}\xspace}
\newcommand{\detajj}{\ensuremath{\abs{\Delta\eta_{\text{jj}}}}\xspace}
\newcommand{\dphijjFAT}{\ensuremath{\abs{\Delta\phi^\text{Wide}_{\text{jj}}}}\xspace}
\newcommand{\detajjFAT}{\ensuremath{\abs{\Delta\eta^\text{Wide}_{\text{jj}}}}\xspace}
\newcommand{\Qstar}{\ensuremath{\PQq^*}\xspace}
\newcommand{\RunLumi}{2.4\fbinv}
\newcommand{\lumiUncert}{12\%\xspace}
\newcommand{\QCDMCscaleFactorDiWideJetMass}{0.88\xspace}
\newcommand{\chiSquare}{31\xspace}
\newcommand{\NDF}{35\xspace}
\newcommand{\jecUncert}{2\%\xspace}
\newcommand{\minMjjCut }{1.2\TeV}
\newcommand{\highestMass }{6.1\TeV}

\date{\today}

\abstract{
A search for narrow resonances in proton-proton collisions
at $\sqrt{s}=13$\TeV is presented.
The invariant mass distribution of the two leading jets is measured with the CMS detector using a data set
corresponding to an integrated luminosity of 2.4\fbinv.
The highest observed dijet mass is 6.1\TeV.
The distribution is smooth and no evidence for resonant particles is observed.
Upper limits at 95\% confidence level are set on the production cross section for narrow resonances with masses
above 1.5\TeV.
When interpreted in the context of specific models, the limits exclude string resonances with masses below
7.0\TeV, scalar diquarks below 6.0\TeV, axigluons and colorons below 5.1\TeV, excited quarks below 5.0\TeV, color-octet
scalars below 3.1\TeV, and $\PWpr$ bosons below 2.6\TeV. These results significantly extend previously published limits.}

\hypersetup{%
pdfauthor={CMS Collaboration},%
pdftitle={Search for narrow resonances decaying to dijets in proton-proton collisions at sqrt(s) = 13 TeV},%
pdfsubject={CMS},%
pdfkeywords={CMS, physics, search, exotica, dijet, resonance}}

\maketitle

Deep inelastic proton-proton ($\Pp\Pp$) collisions often produce two or more energetic jets when the constituent partons are
scattered with large transverse momenta (\pt).
The invariant mass \mjj of the pair of jets having the largest values of \pt in the event (the dijet) has a spectrum that is predicted by quantum chromodynamics (QCD)
to fall steeply and smoothly with increasing dijet mass~\cite{Chatrchyan:2011qta,Aad:2013tea}.
Many extensions of the standard model predict the existence of new massive particles
that couple to quarks (\PQq) and gluons (\Pg) and can be detected as resonances in the
dijet mass spectrum. In this Letter, we report a search for narrow
resonances, those with natural widths that are small compared to the experimental resolution.
The search uses the dijet mass spectrum measured with the CMS detector
at a center-of-mass energy of $\sqrt{s}=13$\TeV. The data
correspond to an integrated luminosity of \RunLumi from the 2015 running of the CERN LHC (Run 2).

The most stringent current bounds on dijet resonance production have been presented by the
CMS~\cite{Khachatryan:2010jd,Chatrchyan2011123,CMS:2012yf,Chatrchyan:2013qhXX,Khachatryan:2015sja} and
\mbox{ATLAS}~\cite{ATLAS2010,Aad:2011aj,Aad201237,ATLAS:2012pu,Aad:2014aqa} Collaborations, using proton-proton
collisions at $\sqrt{s}= 7$ and 8\TeV from LHC Run 1.
A summary of previous searches and a comparison of the different strategies are presented in Ref.~\cite{Harris:2011bh}.
The present search is expected to be more sensitive than previous studies for dijet masses above 2\TeV, and has similar
sensitivity as a contemporaneous search from ATLAS at $\sqrt{s}= 13$\TeV~\cite{ATLAS:2015nsi}.
We present a model-independent search and, in addition, consider the following models of
$s$-channel dijet resonances: string resonances~\cite{Anchordoqui:2008di,Cullen:2000ef}, scalar diquarks~\cite{ref_diquark},  axigluons~\cite{ref_axi,Chivukula:2013xla},
colorons~\cite{ref_coloron,Chivukula:2013xla}, excited quarks
(\Qstar)~\cite{ref_qstar,Baur:1989kv}, color-octet scalars~\cite{Han:2010rf},
new gauge bosons ($\PWpr$ and $\PZpr$)~\cite{ref_gauge}, and Randall--Sundrum (RS) gravitons
(G)~\cite{ref_rsg}. More details on the specific choices of couplings
for these models can be found in Ref.~\cite{CMS:2012yf}.

The CMS detector and its coordinate system, including the azimuthal angle $\phi$ (in radians) and the pseudorapidity $\eta$,
are described in detail in Ref.~\cite{refCMS}.
Events are selected using a two-tier trigger system. Events satisfying
loose jet requirements at the first level (L1) are examined by the
high-level trigger (HLT), where jets are clustered from particle-flow (PF)~\cite{PFT-09-001-PAS,PFT-10-001-PAS}
candidates, discussed in the next paragraph. The jets with $\pt>40$\GeV and $|\eta|<3$ are used to compute $H_\mathrm{T}$, the scalar
sum of the jet \pt. Events are accepted if they have $H_\mathrm{T}>800$\GeV
or include a jet with $\pt>500$\GeV.
At least one reconstructed vertex is required with $\abs{z} <
24\unit{cm}$. The primary vertex is defined as the vertex with the
highest sum of $\pt^2$ of the associated tracks.

The PF algorithm is used to reconstruct the
particles in an event and
to identify them as muons, electrons (with associated Bremsstrahlung
photons), photons (unconverted and converted into $\Pep\Pem$ pairs), and either charged or neutral
hadrons. These PF candidates are clustered into jets using the
anti-$k_t$ algorithm~\cite{antikt} with a distance parameter of 0.4, implemented in the \textsc{FastJet} package~\cite{Cacciari:2005hq}.
Charged PF candidates not originating from the primary vertex
are removed prior to the jet finding.
An event-by-event jet-area-based
correction~\cite{jetarea_fastjet,jetarea_fastjet_pu,Chatrchyan:2011ds}
is applied to the jets to remove the estimated energy from additional collisions in
the same or adjacent bunch crossings (pileup).
The jet momenta and energies are further corrected using calibration constants
obtained from simulation, test beam results, and pp collision
data at $\sqrt{s}=13$\TeV, using methods described in Ref.~\cite{Chatrchyan:2011ds} with all \text{in situ}
calibrations obtained from the current data. All jets are
required to have $\pt>30$\GeV and $\abs{\eta}<2.5$.  The two jets with largest
\pt are defined as the leading jets.
Jet identification criteria~\cite{JME-10-003-PAS} are
applied to remove spurious jets associated with calorimeter
noise. An event is rejected if either of the two leading jets does not satisfy
the jet identification criteria.

Geometrically close jets are combined into ``wide jets'' and
used to determine the dijet mass, as in our previous
searches~\cite{Chatrchyan2011123,CMS:2012yf,Chatrchyan:2013qhXX,Khachatryan:2015sja}.  The wide-jet algorithm, designed 
for dijet resonance event reconstruction, reduces the analysis sensitivity to gluon radiation from the
final state partons.  The two leading jets are used as seeds and the four-vectors of all other jets, if within $\Delta R=\sqrt{\smash[b]{(\Delta\eta)^2 +
  (\Delta\phi)^2}}<1.1$, are added to the
nearest leading jet to obtain two wide jets, which then form
the dijet system. The background from $t$-channel dijet events is
suppressed by requiring the pseudorapidity separation of the two wide
jets to satisfy $\detajj<1.3$.
The above requirements, originally developed for the analysis of Run~1 data, maximize the search sensitivity for
isotropic decays of dijet resonances in the presence of QCD dijet background. It has been verified
that these requirements remain optimal for collisions at $\sqrt{s}=13$\TeV.
We select events with $\mjj>\minMjjCut$ for which the combined L1 trigger and HLT are found to be fully efficient.

\begin{figure}[hbtp]
  \begin{center}
    \includegraphics[width=\cmsFigWidth]{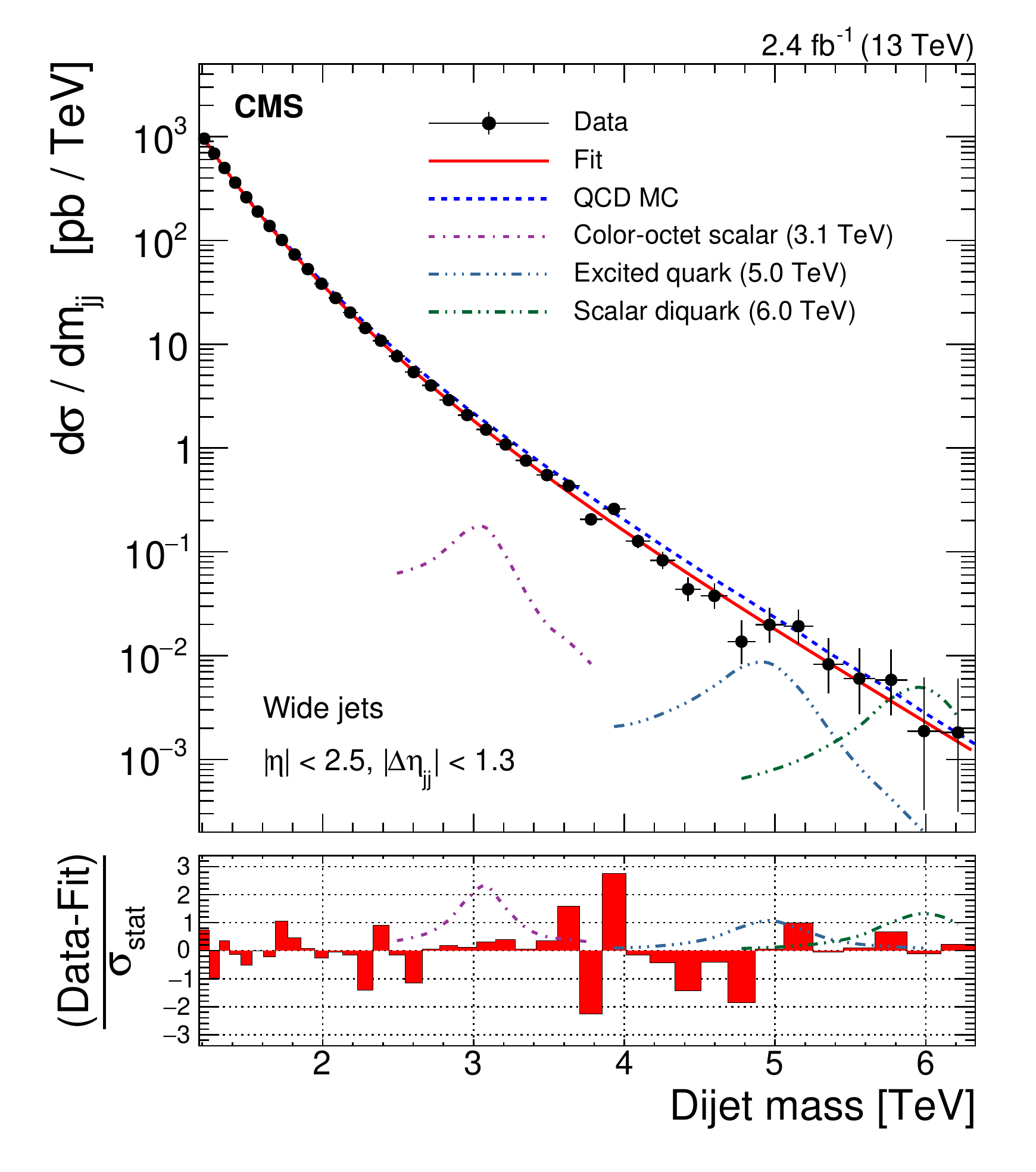}
    \caption{
    Dijet mass spectrum (points) compared to a fitted
  parameterization (solid curve) and to the prediction of
  the {\PYTHIA8}~\cite{Sjostrand:2007gs} QCD MC event generator including
  simulation of the detector (dashed curve).
  The lower panel shows the difference between the data and the
  fitted parametrization, divided by the statistical uncertainties.
    The predicted distributions of narrow resonance signals
  for three models, with resonance mass values corresponding
  to the respective 95\% confidence level exclusion limit, are shown in both panels
  (dash-dotted curves).
}
    \label{figDataAndMC}
  \end{center}
\end{figure}

Figure~\ref{figDataAndMC} shows
the dijet mass spectrum, defined as the observed number of events in each bin divided by the
integrated luminosity and bin width, with predefined bins of width corresponding to the
dijet mass resolution~\cite{Khachatryan:2010jd}.
The highest dijet mass observed is \highestMass.
The data are compared with a leading-order QCD Monte Carlo (MC)
prediction from the {\PYTHIA8} (v205) \cite{Sjostrand:2007gs}
generator with the CUETP8M1 tune~\cite{Khachatryan:2015pea,Skands:2014pea}, including
 a \GEANTfour-based \cite{refGEANT} simulation of the CMS detector.
The \PYTHIA simulation uses the NNPDF2.3LO~\cite{Ball:2012cx} parton distribution functions (PDF).
The renormalization and factorization scales are both set at the \pt value of the hard-scattered
partons. The MC prediction is normalized to the integrated contents
of the data in Fig.~\ref{figDataAndMC}, requiring multiplication
of the predicted cross section by a factor of~\QCDMCscaleFactorDiWideJetMass.

To test the smoothness of the measured dijet mass spectrum, we fit the data with
the parameterization
\begin{equation} \label{eq:FitFunctionA}
\frac{{\rd}\sigma}{{\rd}\mjj} =
\frac{P_{0} (1 - x)^{P_{1}}}{x^{P_{2} + P_{3} \ln{(x)}}}\ ,
\end{equation}
where $x=\mbox{\mjj}/\sqrt{s}$ and $P_0$, $P_1$, $P_2$, and $P_3$ are fitted parameters.
This functional form was also used in previous
searches \cite{refCDFrun2,Khachatryan:2010jd,Chatrchyan2011123,CMS:2012yf,Chatrchyan:2013qhXX,Khachatryan:2015sja,
ATLAS2010,Aad:2011aj,Aad201237,ATLAS:2012pu,Aad:2014aqa} to describe the data and the QCD
predictions. In Fig.~\ref{figDataAndMC}
we show the result of the binned maximum likelihood fit, which yields $\chi^2=\chiSquare$
for \NDF degrees of freedom, where the $\chi^2$ is determined using the Poisson uncertainties shown in Fig.~\ref{figDataAndMC}.
The data are seen to be well described by the fit.

\begin{figure}[hbtp]
  \begin{center}
     \includegraphics[width=\cmsFigWidth]{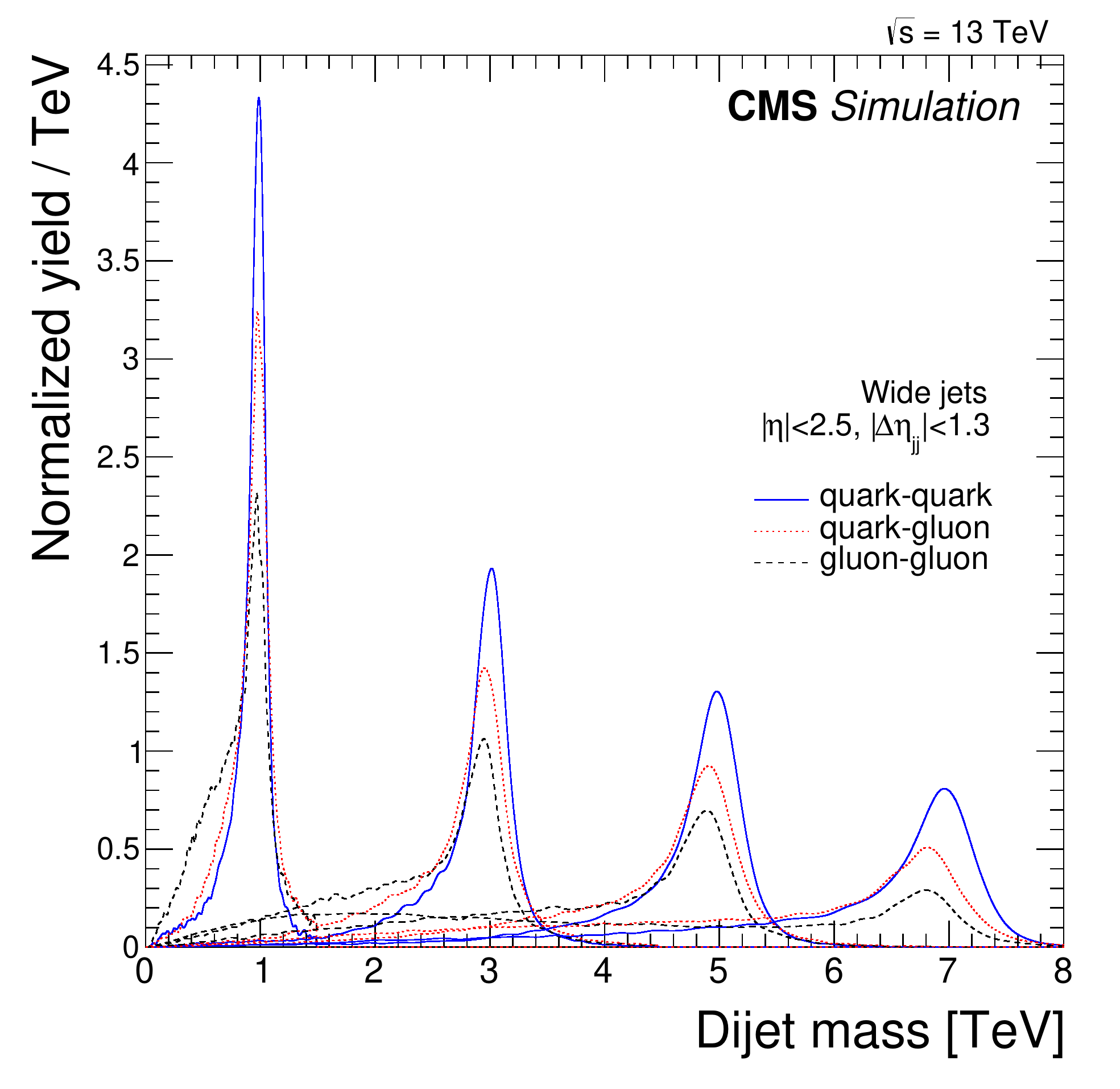}
   \caption{
   The reconstructed resonance mass spectrum predicted by
  the {\PYTHIA8}~\cite{Sjostrand:2007gs} MC event generator, including
  simulation of the detector, for resonances generated with
  mass values 1, 3, 5, and 7\TeV, for quark-quark processes
  modeled by $\PQq\PAQq\to \PXXG \to \PQq\PAQq$ (solid),
for quark-gluon processes modeled by $\PQq\Pg\to \Qstar \to \PQq\Pg$ (dotted),
and for gluon-gluon processes modeled by $\Pg\Pg\to \PXXG \to \Pg\Pg$ (dashed),
where \PXXG is a RS graviton and \Qstar is an excited quark.}
    \label{figMassShapes}
  \end{center}
\end{figure}

We search in the dijet mass spectrum for narrow resonances.
Figure~\ref{figMassShapes} shows example dijet mass distributions for simulated signal events, generated with the \PYTHIA8 program.
The predicted mass distributions have Gaussian cores from the jet energy resolution,
and tails towards lower mass values primarily from QCD radiation. The
contribution of this low-mass tail to the lineshape depends on the
parton content of the resonance ($\PQq\PQq$, $\PQq\Pg$, or $\Pg\Pg$).  Resonances
containing gluons, which emit QCD radiation more strongly than
quarks, have a more pronounced tail.
For the high-mass resonances, there is also a significant contribution that depends both on the
PDF and on the natural width of the Breit--Wigner resonance.  For resonances
produced through interactions of nonvalent partons in the proton, the low-mass component of
the Breit--Wigner resonance distribution is amplified by the rise of the parton probability distribution at
low fractional momentum. These effects cause a large tail at low mass values. Neglecting the tails, the
approximate value of the dijet mass resolution varies with resonance mass from 7\% at 1.5\TeV to 4\% at 7\TeV.

There is no evidence for a narrow resonance in the data, as seen from Fig.~\ref{figDataAndMC}.
The most significant excess in the data relative to the background fit occurs for a dijet mass of 3.9\TeV. A fit to the
hypothesis of a narrow $\PQq\PQq$ resonance, which includes contributions from the bin at 3.9\TeV and neighboring bins, has a local statistical
significance of 1.7 standard deviations.
Figure~\ref{figDataAndMC} includes example signal distributions of the three kinds of narrow resonances ($\PQq\PQq$, $\PQq\Pg$, and $\Pg\Pg$)
at the mass values (6.0, 5.0, and 3.1\TeV) corresponding to the limit set on the respective models
(scalar diquark, excited quark, and color-octet scalar). These limits are presented below.

We use the dijet mass spectrum from wide jets, the background
parameterization, and the dijet resonance shapes to set
limits on new particles decaying to the parton pairs $\PQq\PQq$ (or
 $\PQq\PAQq$), $\PQq\Pg$, and $\Pg\Pg$. A separate limit is determined
for each final state ($\PQq\PQq$, $\PQq\Pg$, and $\Pg\Pg$) because of the dependence of the
dijet resonance shape on the type of the two final-state partons.

The dominant sources of systematic uncertainty are the jet energy scale, jet energy resolution,
integrated luminosity, and the estimation of background. The uncertainty in the jet energy scale is \jecUncert,
determined from Run 2 data using the methods described in Ref.~\cite{Chatrchyan:2011ds}.
This uncertainty is propagated to the limits by shifting the dijet mass for signal events by $\pm$\jecUncert.
The uncertainty in the jet energy
resolution translates into an uncertainty of 10\% in the resolution of the dijet
mass~\cite{Chatrchyan:2011ds}, and is propagated to the limits by increasing and decreasing by 10\% the reconstructed
width of the dijet mass shape for signal.
The luminosity scale and its uncertainty are estimated from
beam-beam scans utilizing the methods from Ref.~\cite{lumiPAS13}. The uncertainty in the
integrated luminosity is \lumiUncert, and is propagated to the normalization of
the signal.
Not included in this analysis is a recent re-evaluation of the integrated luminosity that increases the
measured value by 4.3\% and reduces the systematic uncertainty to 4.6\%.
Changes in the values of the parameters describing the background introduce a change in the signal strength
that is accounted for as a systematic uncertainty.
The dependence of the signal mass distributions on the number of pileup interactions is negligible.

To set upper limits on signal cross sections we use a Bayesian
formalism~\cite{Agashe:2014kda} with a uniform prior for a positive
signal cross section; log-normal priors are used to
model systematic uncertainties in the jet energy scale, jet energy
resolution, and integrated luminosity,
all treated as nuisance parameters to be integrated over. We calculate the likelihood, namely the posterior probability density as a
function of resonance cross section, independently at each value of
resonance pole mass from 1.5 to 7.2\TeV in 0.1\TeV steps.
Resonances with masses less than 1.5\TeV are too close to the lower edge of our dijet mass spectrum to produce
a peak distinguishable from the background and are therefore not considered. Examples of the
resonance shapes used are shown in Fig.~\ref{figMassShapes}.
The data are fitted with the background function plus a signal shape, with the signal cross
section a fitted parameter. The resulting fitting function, with the signal cross
section set to zero, is used as the background estimate.
The likelihood is formed using as input the data, the background estimate from the best fit of the signal+background
hypothesis to the data, and the resonance shape multiplied by the resonance cross section.
The uncertainty in the background is incorporated through marginalization, i.e., by
integrating the likelihood over the background parameters using uniform priors.
The integration is performed for each of the background nuisance parameters
in a range around the best-fit values, corresponding to a decrease in the
likelihood by a factor of 1000 from its maximum value for each parameter
independently.

\begin{figure*}[hbtp]
  \centering
    \includegraphics[width=0.48\textwidth]{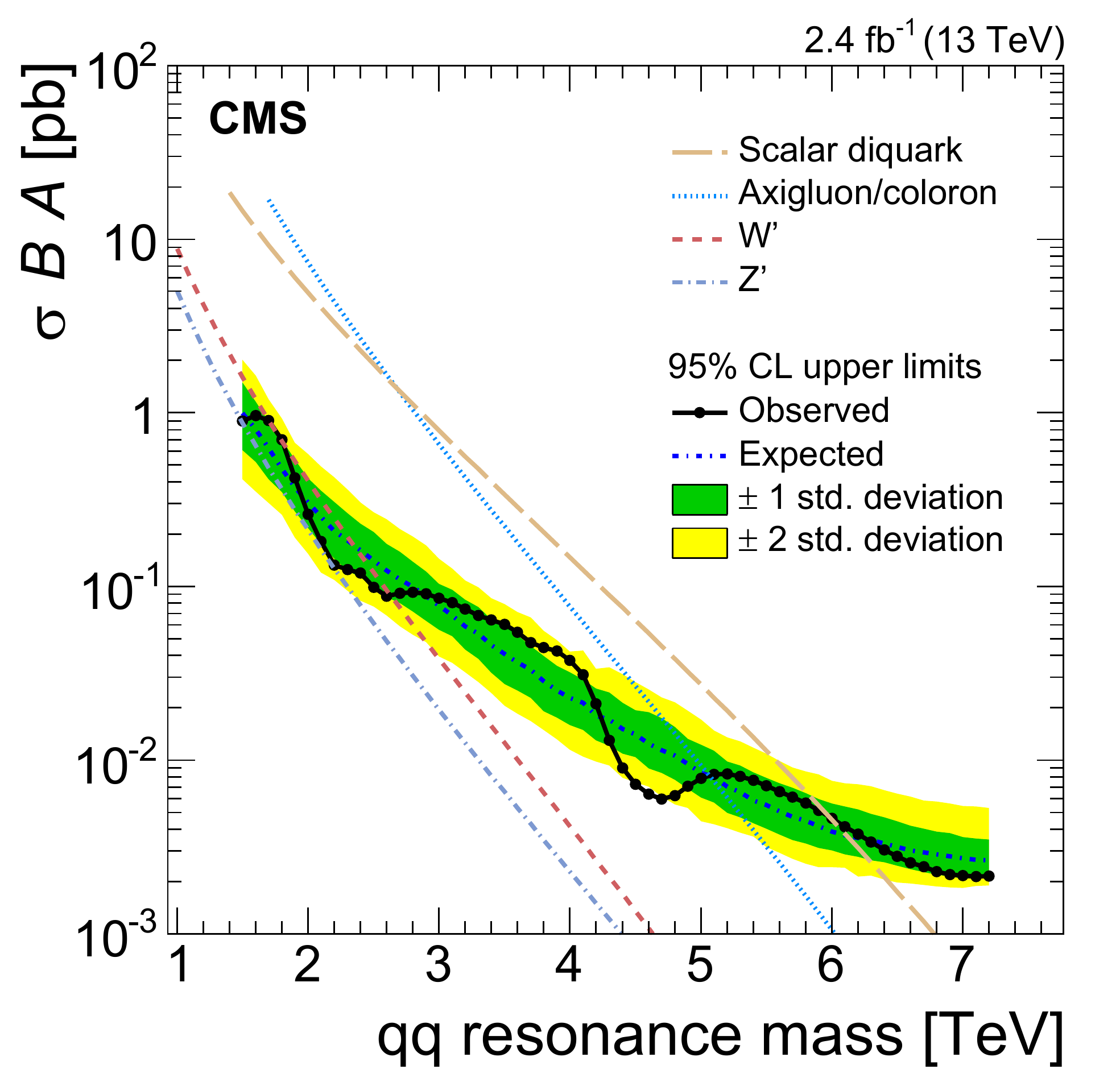}
    \includegraphics[width=0.48\textwidth]{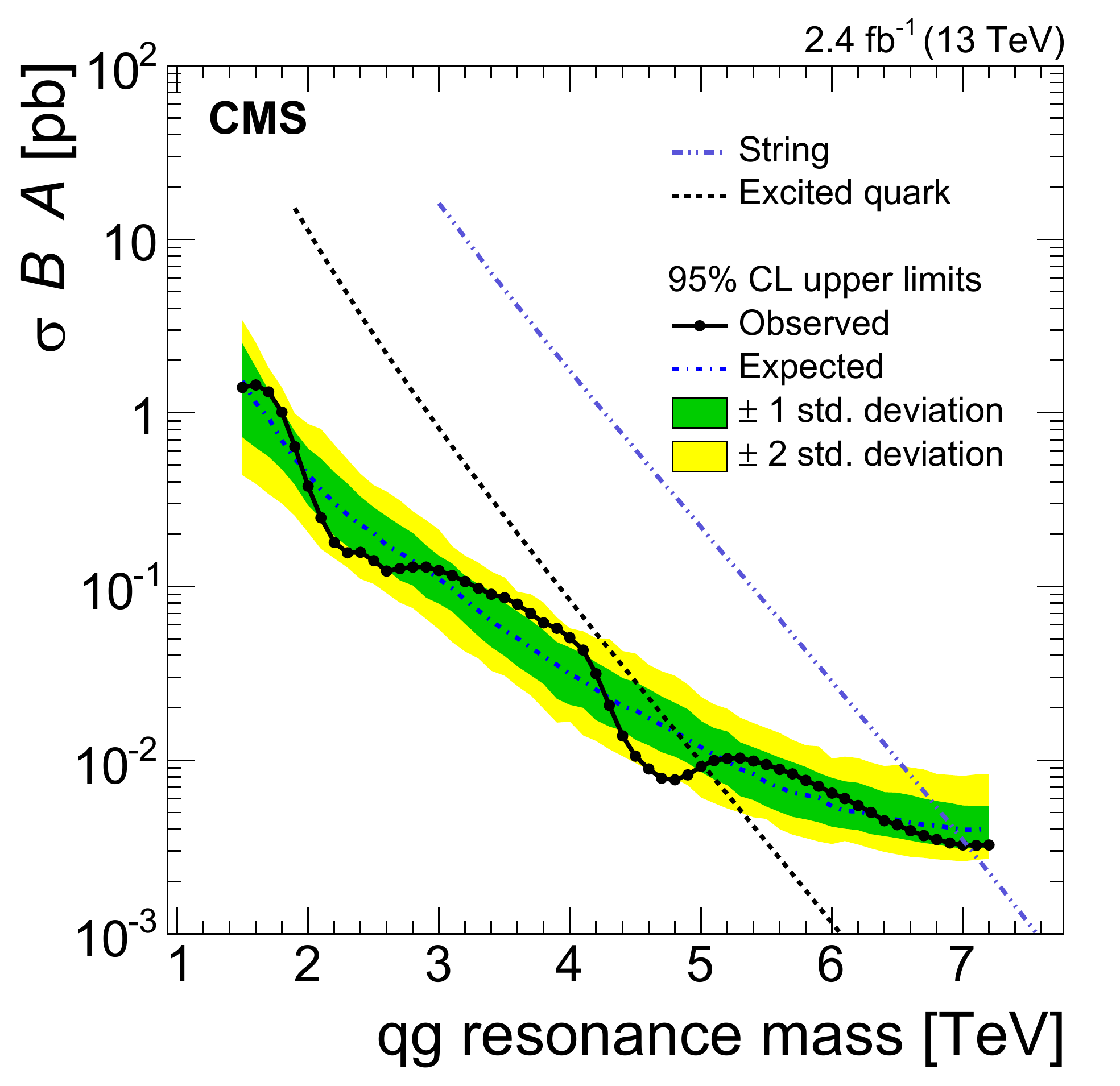}
    \includegraphics[width=0.48\textwidth]{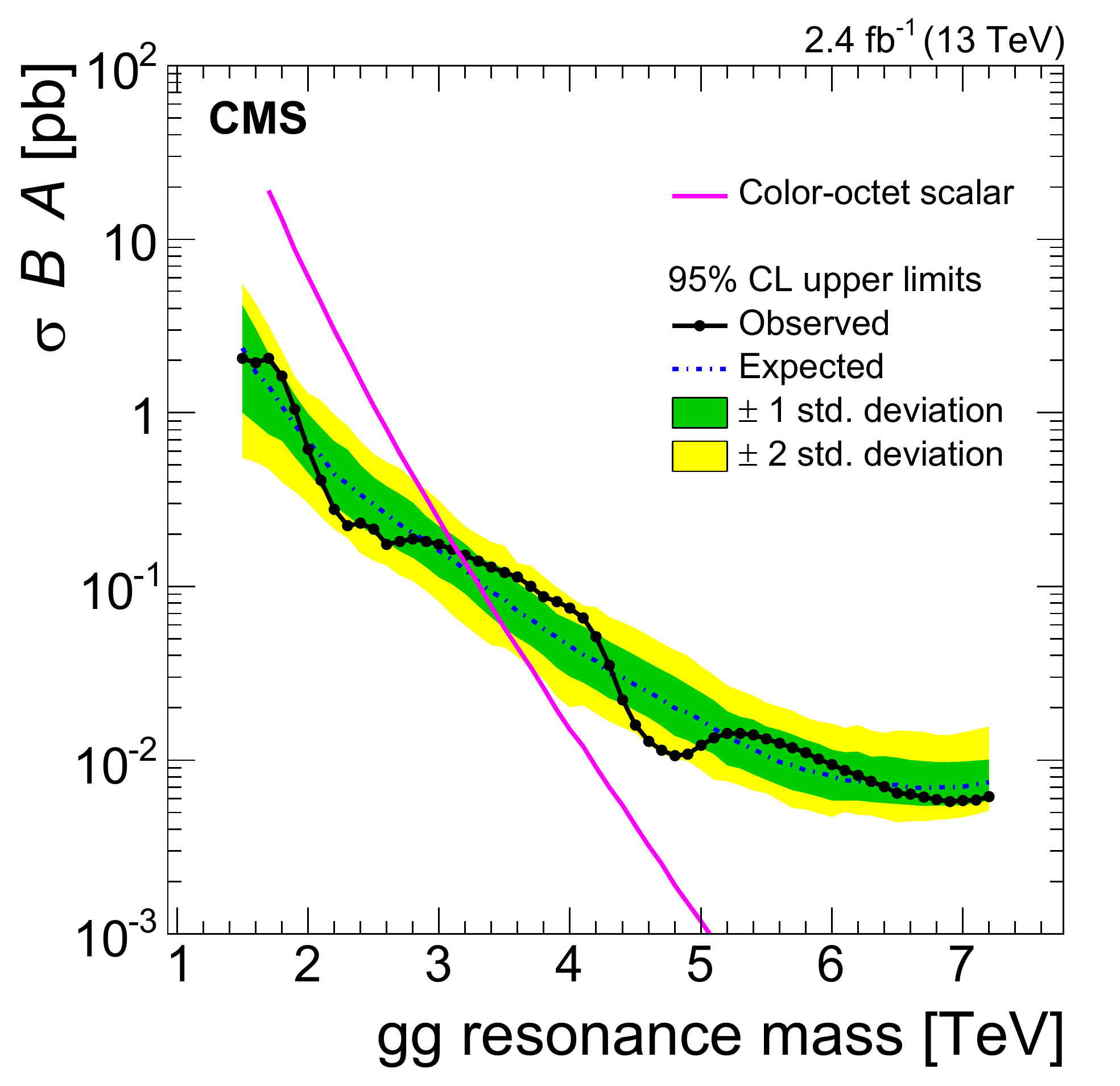}
    \includegraphics[width=0.48\textwidth]{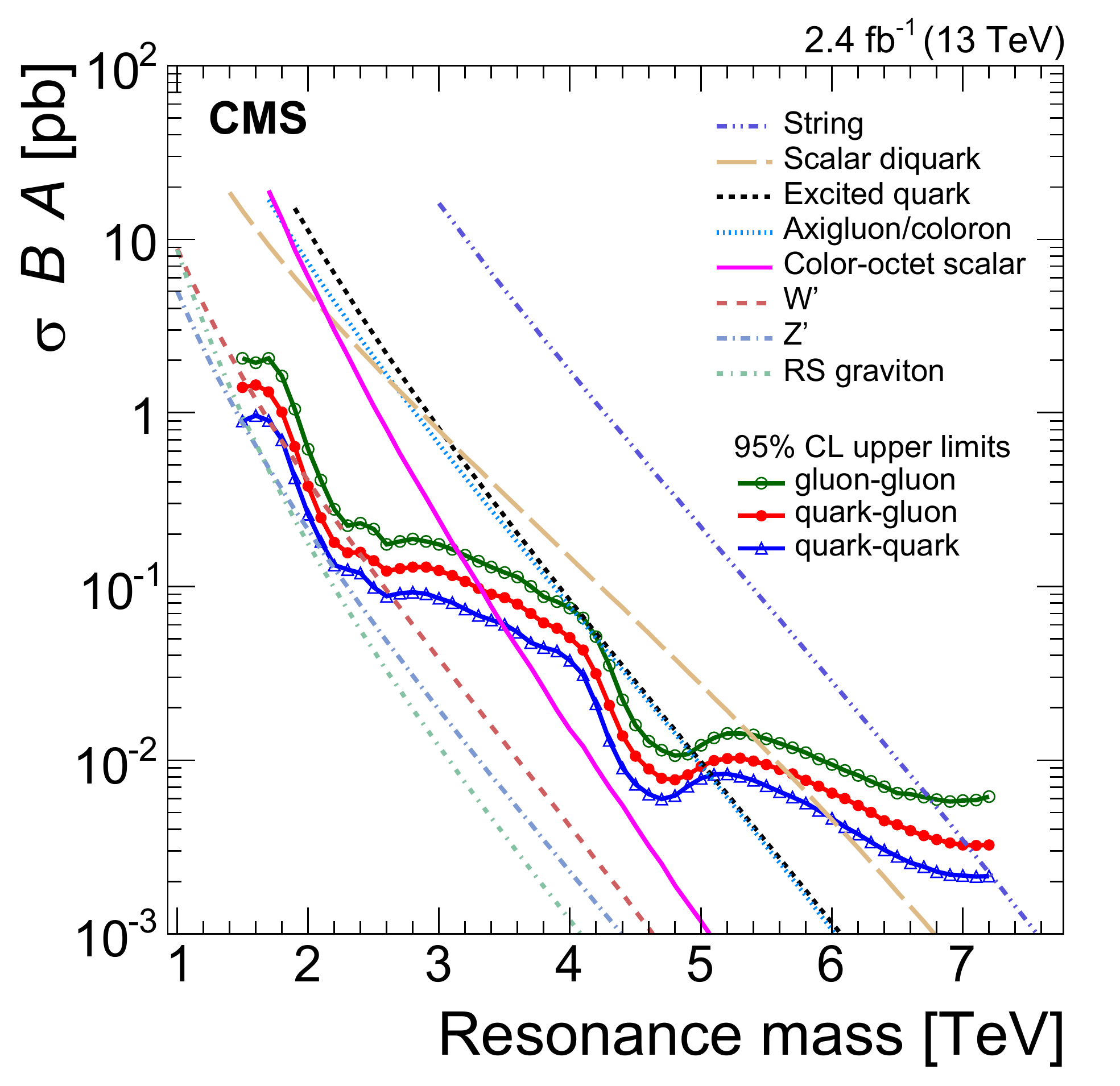}
    \caption{The observed 95\% \CL upper limits on the product of the cross section, branching fraction, and acceptance for
    quark-quark (top left), quark-gluon (top right),
    and gluon-gluon (bottom left) type dijet resonances, shown as symbols connected by solid curves, and a comparison of all types
    (bottom right). The corresponding expected limits (dash-dotted curves) and their variation at the 1 and
    2 standard deviation levels (shaded bands) are also shown.  The limits are compared to the predicted cross sections of  string resonances~\cite{Anchordoqui:2008di,Cullen:2000ef}, scalar diquarks~\cite{ref_diquark},
excited quarks~\cite{ref_qstar,Baur:1989kv}, axigluons~\cite{ref_axi,Chivukula:2013xla},
colorons~\cite{ref_coloron,Chivukula:2013xla}, color-octet scalars~\cite{Han:2010rf},
new gauge bosons $\PWpr$ and $\PZpr$~\cite{ref_gauge},
and RS gravitons~\cite{ref_rsg}.}
    \label{figLimit}
\end{figure*}
Figure~\ref{figLimit} shows the model-independent observed
upper limits at 95\% confidence level (\CLp) on
$\sigma\, B\, A$, i.e., the product of
the cross section ($\sigma$), the branching fraction ($B$), and the acceptance ($A$) for the
kinematic requirements $\detajj<1.3$ and $\abs{\eta}<2.5$,
for narrow $\PQq\PQq$, $\PQq\Pg$, and $\Pg\Pg$ resonances.
The acceptance of the requirement $\mjj>\minMjjCut$ has been taken into account
by correcting the limits, and therefore does not appear in the acceptance $A$.
Figure~\ref{figLimit} also shows the expected limits on the cross section and their bands of uncertainty.
The expected limits are estimated with pseudo-experiments generated
using background-only hypotheses. The generated mass spectra are
fit with a background+signal model to extract expected upper limits.
The difference in the limits for $\PQq\PQq$, $\PQq\Pg$, and $\Pg\Pg$ resonances at the same resonance mass
originates from the difference in their lineshapes.
All upper limits presented are compared to the parton-level predictions of $\sigma\, B\, A$, without detector simulation,
to determine mass limits on new particles.
The model predictions shown in Fig.~\ref{figLimit} are calculated in the narrow-width
approximation~\cite{Harris:2011bh} using the CTEQ6L1~\cite{refCTEQ} PDF at leading order,
with a next-to-leading order correction factor of approximately 1.3 included for the $\PWpr$ and $\PZpr$
models~\cite{Gumus:962025}, and approximately 1.2 for the axigluon and coloron models~\cite{Chivukula:2013xla}.
The branching fraction B includes the direct decays of the resonance into $\PQq\PQq$, $\PQq\Pg$, and $\Pg\Pg$ only, excluding top quarks,
and no other decays giving jets are considered in the theoretical calculation of the model cross section.
The acceptance is evaluated at the parton level for the resonance decay to two partons. In the case of isotropic
decays it is $A\approx 0.6$ independent of resonance mass.

For a given model, new particles are excluded at 95\% \CL in mass regions where the theoretical prediction
lies at or above the observed upper limit for the appropriate final state of Fig.~\ref{figLimit}.
The mass limits reported in Table~\ref{tab:MassLimit}
 represent significant extensions of the most stringent observed limits from LHC Run~1~\cite{Khachatryan:2015sja,Aad:2014aqa}.
For string resonances, the observed mass limit of 7.0\TeV extends the previous CMS limit of 5.0\TeV;
for scalar diquarks, the observed mass limit of 6.0\TeV extends the previous CMS limit of 4.7\TeV;
for axigluons and colorons, the observed mass limit of 5.1\TeV extends the previous CMS limit of 3.6\TeV;
for excited quarks, we set a mass limit of 5.0\TeV compared to the ATLAS limit of 4.06\TeV;
for a color-octet scalar, the observed mass limit of 3.1\TeV improves the ATLAS limit of 2.70\TeV;
and for a $\PWpr$ boson, we exclude masses up to 2.6\TeV, just beyond the ATLAS limit of 2.45\TeV.
With the current data sample we cannot set mass limits on $\PZpr$
bosons with standard-model-like couplings or on RS gravitons with dimensionless coupling less than 0.1.

In summary, a search for narrow resonances decaying into a pair of jets has been performed using
a data sample of pp collisions at $\sqrt{s}=13$\TeV corresponding to
an integrated luminosity of \RunLumi.
The dijet mass spectrum has been measured to be a smoothly falling
distribution. In the analyzed data sample, there is no evidence for resonant particle production.
We present generic upper limits on the product $\sigma\, B\, A$ that are applicable to any model
of narrow dijet resonance production.
This search is more sensitive than previous searches for dijet
resonances for masses above 2\TeV.
We set the most stringent limits to date on the masses of string resonances, scalar diquarks, axigluons, colorons, excited quarks,
color-octet scalars, and $\PWpr$ bosons.

\begin{table}[ht]
  \caption{Observed and expected mass limits for analyses that
  exclude the listed models at 95\%~\CL for a resonance mass from 1.5\TeV up to the indicated values.}
  \centering
  \begin{scotch}{lccc}
    Model & Final & Obs. Mass & Exp. Mass \\
          &  State           & Limit [\TeVns{}]     & Limit [\TeVns{}] \\
    \hline
String   & $\PQq\Pg$   &  7.0  & 6.9 \\
Scalar diquark  & $\PQq\PQq$  & 6.0 & 6.1 \\
Axigluon/coloron  & $\PQq\PAQq$  & 5.1  & 5.1 \\
Excited quark (\Qstar) &  $\PQq\Pg$ & 5.0  & 4.8 \\
Color-octet scalar  & $\Pg\Pg$  & 3.1 & 3.3 \\
Heavy $\PW$ ($\PWpr$) & $\PQq\PAQq$ & 2.6 & 2.3 \\
  \end{scotch}
\label{tab:MassLimit}
\end{table}

\begin{acknowledgments}
We congratulate our colleagues in the CERN accelerator departments for the excellent performance of the LHC and thank the technical and administrative staffs at CERN and at other CMS institutes for their contributions to the success of the CMS effort. In addition, we gratefully acknowledge the computing centers and personnel of the Worldwide LHC Computing Grid for delivering so effectively the computing infrastructure essential to our analyses. Finally, we acknowledge the enduring support for the construction and operation of the LHC and the CMS detector provided by the following funding agencies: BMWFW and FWF (Austria); FNRS and FWO (Belgium); CNPq, CAPES, FAPERJ, and FAPESP (Brazil); MES (Bulgaria); CERN; CAS, MoST, and NSFC (China); COLCIENCIAS (Colombia); MSES and CSF (Croatia); RPF (Cyprus); MoER, ERC IUT and ERDF (Estonia); Academy of Finland, MEC, and HIP (Finland); CEA and CNRS/IN2P3 (France); BMBF, DFG, and HGF (Germany); GSRT (Greece); OTKA and NIH (Hungary); DAE and DST (India); IPM (Iran); SFI (Ireland); INFN (Italy); MSIP and NRF (Republic of Korea); LAS (Lithuania); MOE and UM (Malaysia); CINVESTAV, CONACYT, SEP, and UASLP-FAI (Mexico); MBIE (New Zealand); PAEC (Pakistan); MSHE and NSC (Poland); FCT (Portugal); JINR (Dubna); MON, RosAtom, RAS and RFBR (Russia); MESTD (Serbia); SEIDI and CPAN (Spain); Swiss Funding Agencies (Switzerland); MST (Taipei); ThEPCenter, IPST, STAR and NSTDA (Thailand); TUBITAK and TAEK (Turkey); NASU and SFFR (Ukraine); STFC (United Kingdom); DOE and NSF (USA).
\end{acknowledgments}

\bibliography{auto_generated}
\cleardoublepage \appendix\section{The CMS Collaboration \label{app:collab}}\begin{sloppypar}\hyphenpenalty=5000\widowpenalty=500\clubpenalty=5000\textbf{Yerevan Physics Institute,  Yerevan,  Armenia}\\*[0pt]
V.~Khachatryan, A.M.~Sirunyan, A.~Tumasyan
\vskip\cmsinstskip
\textbf{Institut f\"{u}r Hochenergiephysik der OeAW,  Wien,  Austria}\\*[0pt]
W.~Adam, E.~Asilar, T.~Bergauer, J.~Brandstetter, E.~Brondolin, M.~Dragicevic, J.~Er\"{o}, M.~Flechl, M.~Friedl, R.~Fr\"{u}hwirth\cmsAuthorMark{1}, V.M.~Ghete, C.~Hartl, N.~H\"{o}rmann, J.~Hrubec, M.~Jeitler\cmsAuthorMark{1}, V.~Kn\"{u}nz, A.~K\"{o}nig, M.~Krammer\cmsAuthorMark{1}, I.~Kr\"{a}tschmer, D.~Liko, T.~Matsushita, I.~Mikulec, D.~Rabady\cmsAuthorMark{2}, B.~Rahbaran, H.~Rohringer, J.~Schieck\cmsAuthorMark{1}, R.~Sch\"{o}fbeck, J.~Strauss, W.~Treberer-Treberspurg, W.~Waltenberger, C.-E.~Wulz\cmsAuthorMark{1}
% ...
\vskip\cmsinstskip
% ...
1:~~Also at Vienna University of Technology, Vienna, Austria\\
2:~~Also at CERN, European Organization for Nuclear Research, Geneva, Switzerland\\
% ...

\end{sloppypar}
\end{document}